\documentclass[%
 pre,twocolumn,
 amsmath,amssymb,
 aps,
]{revtex4-2}
\usepackage{xcolor}
\usepackage{graphicx}% Include figure files
\usepackage{dcolumn}% Align table columns on decimal point
\usepackage{bm}
\usepackage{bbold}% bold math
\usepackage{tabularx, multirow,booktabs}
\usepackage{soul}

\begin{document}

%\preprint{APS/123-QED}

\title{Defects in vibrated monolayers of equilateral triangular prisms}

\author{Jorge Vega}
\affiliation{Departamento de F\'{\i}sica Te\'orica de la Materia Condensada,  
Universidad Autónoma de Madrid E-28049 Madrid, Spain}

\author{Enrique Velasco}
 \email{enrique.velasco@uam.es}
\affiliation{Departamento de F\'{\i}sica Te\'orica de la Materia Condensada, Instituto de la 
Materia Condensada (IFIMAC) and Instituto de Ciencia de Materiales Nicol\'as Cabrera, 
Universidad Autónoma de Madrid E-28049 Madrid, Spain}

\author{Yuri Martínez-Ratón}
 \email{yuri@math.uc3m.es}
\affiliation{Grupo Interdisciplinar de Sistemas Complejos (GISC), Departamento de 
Matem\'aticas, Escuela Polit\'ecnica Superior, Universidad Carlos III de Madrid, Avenida de la Universidad 30, E-28911 Legan\'es, Madrid, Spain}

\date{\today}

\begin{abstract}
We present experimental results in which a quasimonolayer of grains, shaped as equilateral triangular prisms, is vertically vibrated within circular and square confining cavities. The system exhibits a fluid phase characterized by sixfold orientational order. However, the specific geometries of the cavities frustrate this order, leading to the excitation of topological defects that adhere to topological principles. We analyze the distribution of topological charges of these defects and their evolution as a function of the container geometry.
In particular, we find that in the circular cavity, both the number and charge of defects remain constant in the steady-state regime, with only defects of charge $+1$ being excited. In contrast, within the square cavity --with the same topology as the circular cavity--, some defects become pinned to the corners of the cavity, developing charges of $+1$ or $+2$. Additionally, free defects with charges of $\pm 1$ can be excited. Notably, dynamic events occur, such as the fusion of two defects or the creation of two defects from a single one. These events are governed by the conservation of total charge within the cavity and give rise to various system configurations, with rapid transitions between them.
The observed dynamic fluctuations in topological charge are remarkable, influenced by both particle shape and cavity geometry. This system of vertically vibrated (dissipative) granular monolayers presents a unique and simple platform for studying topological phenomena, with potential implications for topological effects in equilibrium-oriented fluids.
\end{abstract}

\maketitle

%\tableofcontents

\section{\label{introduction}Introduction}
Vertically-vibrated monolayers of granular particles played an important role in the understanding of
crystallization \cite{Kudrolli,Urbach,Aranson1,Aranson2,Soto}. But recently the tendency of anisotropic grains to form liquid-crystalline phases has
been emphasized \cite{Menon1}. These dissipative monolayers behave in some respects as their thermal-equilibrium counterparts
\cite{Galanis1,Galanis2,Menon1,Menon2,Mexicanos,Daniel,Nosotros1,Nosotros2}. The liquid-crystalline phase diagram of quasimonolayers of cylinders (projecting
as hard rectangles on the plane) has been mapped out as a function of density and particle aspect ratio \cite{Daniel}. 
The diagram shows many similarities with that of equilibrium hard rectangular particles \cite{Dertli}. More recently,
experiments on hard rectangles have shown the excitation of defects in a circular cavity, with their number and topological charge exactly as predicted by topology \cite{Nosotros3}. The four defects behave as repulsive particles with interactions mediated by the orientational stiffness of the four-fold symetric, 4-atic liquid-crystal phase. The structure of the nematic field in the neighbourhood of the defects indicates that they have a topological charge $+1$, giving a total charge of $+4$ within the cavity, as predicted by topology \cite{Lavrentovich,Bowick}. Recent experimental and theoretical studies on colloidal monolayers of hard discorectangles in annular confinement have also shown the formation of smectic textures with different competing defect structures \cite{Wittmann}.

Recent theoretical investigations, based on 
Landau de-Gennes theory, density functional theory, and MC simulations, have examined the influence of geometrical confinement, particularly corners in cavities and protruding obstacles, on the formation and stability of topological defects in systems of rod-like particles \cite{Robinson,Han,Yao}. In the bulk, such systems exhibit 2-atic (uniaxial nematic) order, which supports defects with winding numbers of $k=\pm 1/2$. 

It is interesting to explore whether these findings, pertaining to systems exhibiting 2-atic or 4-atic symmetries, can be extended to other particle shapes, whose intrinsic ordering gives rise to higher-order symmetries in the fluidized granular monolayer. In such cases, confinement within cavities of varying geometries may similarly determine the number and charge of the topological defects through symmetry-based topological arguments. 

The present work considers a system composed of hard equilateral triangles, which forms a $6$-atic phase in bulk \cite{Zhao,Dijkstra,Nosotros4}. Accordingly, the relevant topological defects in this system are characterized by winding numbers of $\pm 1/6$, which reflect the sixfold rotational symmetry of the underlying order. While previous studies typically quantify defect topology using the winding number $k$, and apply conservation laws by summing over all winding numbers in the cavity, resulting from the presence of confining corners or obstacles, we adopt a complementary formulation. Specifically, we use the topological charge of a defect, $q=pk$, where $p$ denotes the symmetry order of the $p$-atic phase ($p=6$ in our case). This approach provides a direct link between defect topology and the underlying symmetry of the ordered phase. The two frameworks, one based on winding numbers and the other on topological charges, are mathematically equivalent in terms of the conservation laws that arise from the topological constraints imposed by corners and obstacles.

In this paper  we explore the behavior of grains with the shape of an equilateral triangular prism, confined into cavities of different geometries. As mentioned before, these particles exhibit an oriented fluid phase with sixfold symmetry (6-atic phase) in equilibrium \cite{Zhao,Dijkstra,Nosotros4}. We show that granular monolayers made of these particles also show these symmetries. However, their behavior 
is greatly affected by the geometry of the container. We have explored two types of cavities: circular and square, both topologically equivalent. Circular and square cavities are both incompatible with the inherent symmetry of the two-dimensional (2D) fluid, and consequently defects are expected to be excited in the system, with a total topological charge given by the genus of the cavity, the total charge being equal to $+6$. We will see that this holds for equilateral triangles, where six defects with charge $+1$ are formed in the circular cavity. By contrast, the square cavity may generate from four to up to seven defects of different charges, while still maintaining the overall charge. The distribution and dynamics of the charges in this case are rather peculiar, with different configurations and number of defects dynamically changing in time. 

Vertically-vibrated granular monolayers are a convenient experimental system to explore the interplay between liquid-crystal symmetries and boundary geometry, which is governed by the rules of topology. Access to different types of defects and their dynamics can be easily tuned to create geometric frustration of various sorts. Equilibrium elastic theory and topological considerations for liquid-crystalline systems seem to explain the phenomenology of the granular monolayers, at least under reasonable assumptions. Also, the effect of dissipation and nonequilibrium phenomena can be assessed by conducting thermal-equilibrium simulations of similar systems in 2D; possibly different behaviours can be inferred from a comparison between the two systems.

The article is organized as follows. In Section \ref{theory} we introduce the elastic theory for
a two-dimensional liquid crystal and the structure of possible defects found in the system. These
defects should abide by the rules of topology in systems with inherent orientational order. Section
\ref{experimental} presents the experimental setup and gives some details on the Monte Carlo (MC) simulations and the protocols used to prepare the systems. Results are presented in Section
\ref{results}, which also contains a detailed discussion of the findings. Section \ref{conclusions} presents some conclusions. Finally, in the Appendix we generalize the topological rules governing the charge and number of defects arising from the confinement of triangular prisms within circular and square cavities to cavities of any regular polygonal shape.

\section{\label{theory}Elastic theory and topological defects}

The concept of geometric frustration describes a
situation where some kind of local order cannot propagate to the entire system. 
Geometric frustration can be caused by appropriate boundary conditions imposed to the system, 
as it occurs in our experiment. Indeed, the director field defined by the
orientation of anisotropic particles in 2D is a
perfect example of a system that can develop geometric
frustration. In our particular case, the sixfold symmetry
of the 6-atic phase is incompatible with both the circular and the 
square symmetry of the boundary. The 6-atic symmetry can be expressed locally by the invariance of the angular distribution function $f(\phi)$ with respect to rotations
of $\pi/3$, i.e. $f(\phi)=f(\phi+\pi/3)$, $f(\phi)$ being the angular probability density of a particle axis forming an angle $\phi$ with respect to the local nematic director. 
In the present case the particle axis of equilateral triangles (the cross-sections of the triangular prisms) points from the barycenter to one of its vertices. Obviously there are three equivalent axes.

\begin{figure*}
\includegraphics[width=0.65\textwidth]{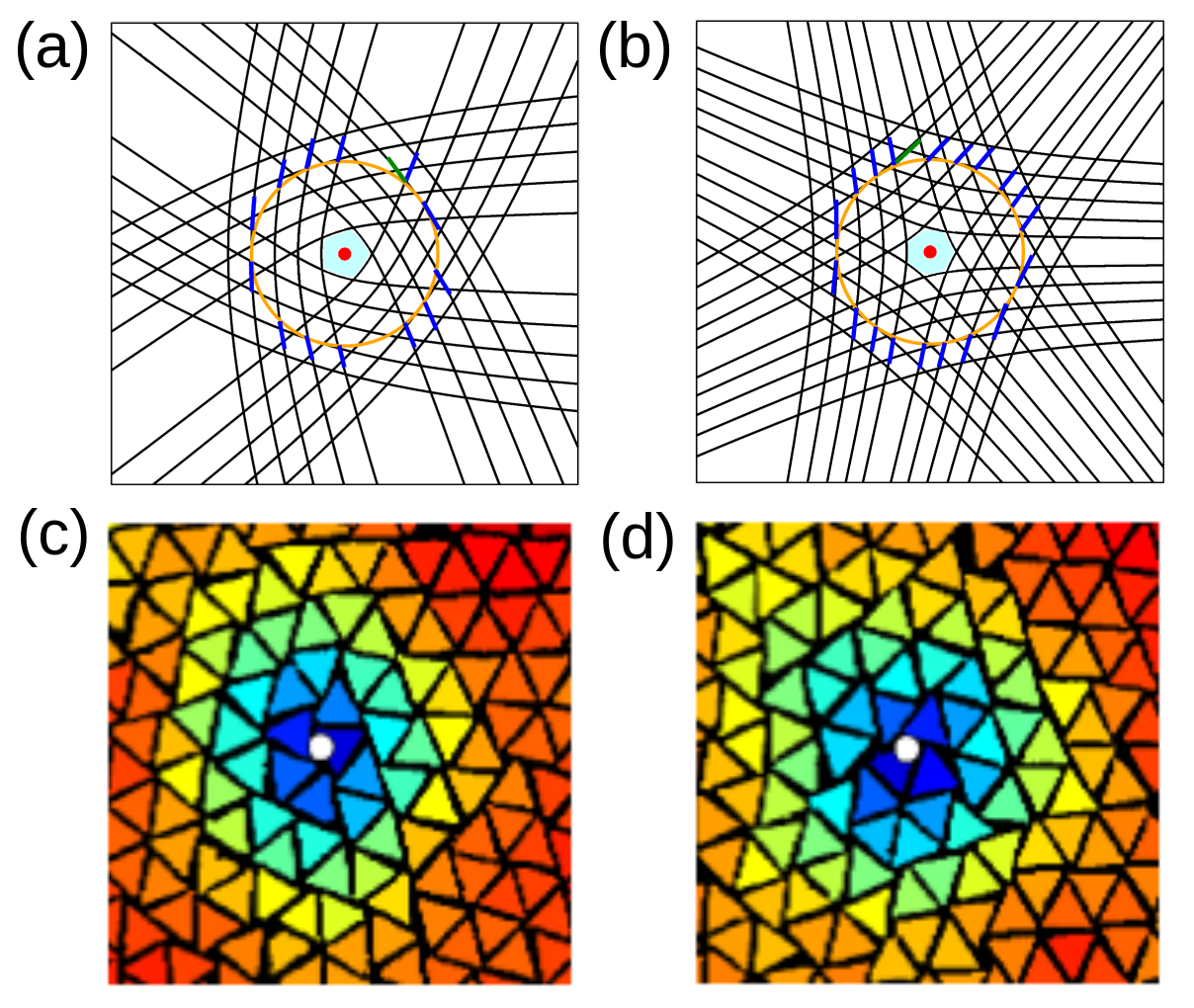}
\caption{\label{fig1} (a) Director field in the neighborhood of   
free defects (filled circle) of the $6$-atic orientational field corresponding 
to a topological charge of  $+1$. (b) Same as (a), but for a charge $-1$. Blue segments in both panels 
indicate the rotation of the director along the circumference shown. 
The total rotation angles are $\pi/3$ for panel (a), and $-\pi/3$ for panel (b). Defect core geometries are shaded. (c) Particle configurations of the granular monolayer in a region identified as a defect of charge $+1$. (d) Same as (c), but for a defect of charge $-1$. Note the close similarities between the structures in (a) and (c), and in (b) and (d). 
}
\end{figure*}

In order to understand how such a system can manifest geometric frustration due to the boundary conditions, we need to invoke some topology concepts based on the Euler theorem of topology \cite{Bowick}. Briefly, this theorem
relates the number of vertices $V$, edges $E$ and faces $F$ of a tessellation of an orientable closed 2D surface as follows:
\begin{eqnarray}
V - E + F = \chi = 2(1 - g) - h
\end{eqnarray}
where $\chi$ is the Euler characteristic of the surface, which
can be expressed in terms of two integers, $g$ the genus (representing the number of `handles'
of the surface), and $h$ (the number of
boundaries or `holes' of the surface). The theorem can be rephrased \cite{Bowick} to show that the total topological charge $q_{\rm t}$ of a system satisfies the following equation:
\begin{eqnarray}
q_{\rm t} =\sum_{i=1}^N q_i=p\chi,
\end{eqnarray}
where $\{q_i\}$ are the individual topological charges of the 
defects and $p$ is the symmetry of the director field (the local orientation is defined modulo 
$2\pi/p$). The ratio between the charge of a defect $q$
and the symmetry of the field $p$ is called the winding
number of the defect $k = q/p$. For example, a simple
vector field ($p = 1$) in a sphere (no handles or holes, so $\chi = 2$) can have two topological defects with charge $q = 1$, that is one source and one sink, a case corresponding to the famous `hairy ball' theorem \cite{Milnor}. 
In our case, the director field of the 6-atic phase has sixfold symmetry 
($p = 6$) and both the circular and square cavities are topologically equivalent
to a disc (without handles but with one boundary, giving $\chi=1$), and therefore $q_{\rm t} = +6$.
Note that Euler's theorem only determines the total topological charge in the system, not the
number of defects or their individual charges.

\begin{figure*}
\includegraphics[width=0.65\textwidth]{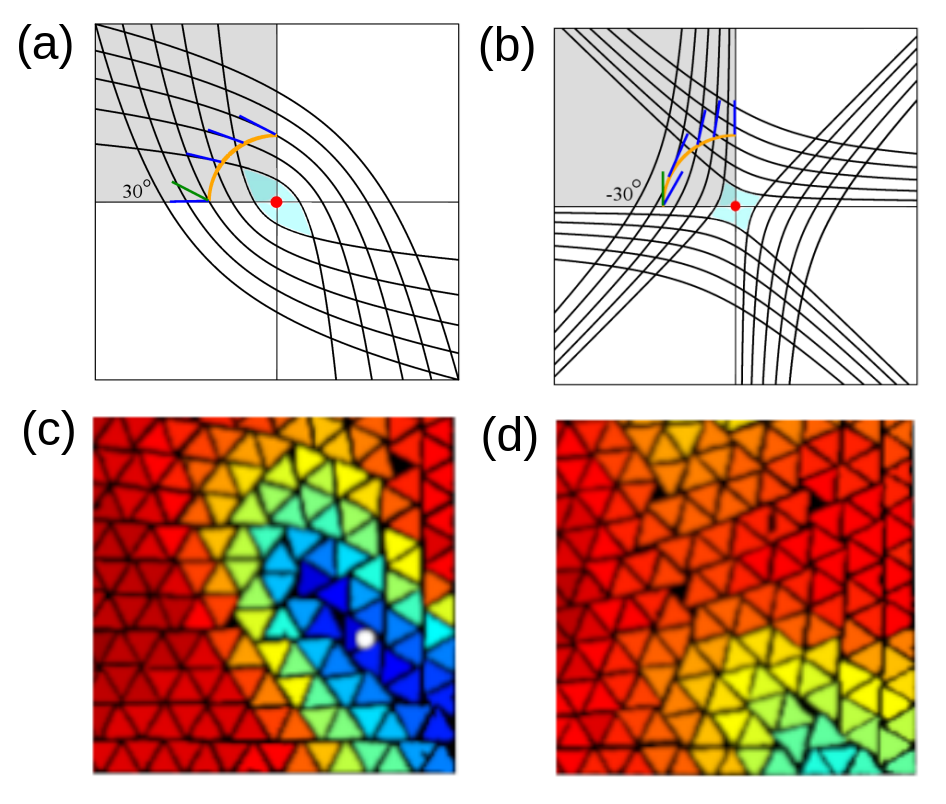}
\caption{\label{fig2} Director fields in the neighborhood of pinned defects (filled circle) of charge (a) $q=+2$, and (b) $q=+1$. Relevant structures of the fields are those inside the shaded quadrants. The total topological charge in each case can be calculated as $q=(\theta+\pi/2)/(2\pi/p)$, with $\theta=\pi/6$
for the $q=+2$ charge and $\theta=-\pi/6$ for the $q=+1$ charge, with $p=6$ as corresponds to a 6-atic symmetry. Rotation angles $\theta=\pm \pi/6$ 
can be obtained by imposing free defect structures with a
global rotation angle of $\pm 2\pi/3$ (i.e. with effective winding numbers $\tilde{k}=\pm 1/3$).
(c) Particle configurations of the granular monolayer in a corner region identified as a defect of charge $+2$. (d) Same as (c), but for a defect of charge $+1$. Note the close similarities between the structures in (a) and (c), and in (b) and (d).}
\end{figure*}

The nematic director field in the neighborhood of a single defect is different according to the topological
charge of the defect. In this work we use the director field in our granular samples to identify the value of the
charges. The relation between charge and director field is obtained from Frank's elastic 
theory \cite{Stewart}, which expresses the cost in free energy associated to a spatial deformation of the
local nematic director ${\bm n}({\bm r})$ as
\begin{eqnarray}
        {\cal F}_{\rm el}[{\bm n}]=\frac{1}{2}\int_A d{\bm r} 
                \left[K_1 \left(\text{div}\  {\bm n}\right)^2 +
        K_3 \left|{\bm n}\times \text{rot} \ {\bm n}\right|^2\right], 
\end{eqnarray}
with $K_1$ and $K_3$ the elastic constants for splay and bend deformations, 
respectively. Considering 2D nematics, we can write
${\bm n}=\left(\cos \theta,\sin\theta\right)$ with
$\theta$ the angle of the nematic director with respect to a fixed laboratory frame.
Assuming the one-constant approximation ($K_1=K_3\equiv K$ which is exact only for 4-atic ordering), the elastic free energy
for a 2D $p$-atic phase is
\begin{eqnarray}
        {\cal F}_{\rm el}[\theta]=\frac{K}{2}\int_A d{\bm r} \left|\boldsymbol{\nabla} \theta\right|^2.
\end{eqnarray}
Functional minimization with respect to $\theta$ provides the Euler-Lagrange equation $\nabla^2\theta=0$,
which generates harmonic solutions. 
Assuming a defect of winding number $k$ at the origin, which represents a point charge, the
constraint $\int_{\gamma} d\theta=2\pi k$ must be satisfied (meaning that the  closed line integral 
along a circle of unit radius $\gamma$, surrounding the defect, is equal to $2\pi k$).
Solutions of the equation satisfying the constraint are given by 
$\theta=k\varphi+\theta_0$, where $\varphi$ is the angle in polar coordinates and $\theta_0$ is a phase. The equilibrium free-energy is then
\begin{eqnarray}
        {\cal F}_{\rm el}=\pi Kk^2\log\left(\frac{R}{a}\right),
\end{eqnarray}
with $a$ the radius of the core of the defect and $R$ the system length. In case of having $N$ defects of the same core size $a$, the associated elastic free energy can 
be evaluated similarly \cite{Chaikin} to give
\begin{eqnarray}
	\frac{{\cal F}_{\rm el}}{\pi K}=
\left[\left(\sum_{i=1}^N k_i\right)^2\log\left(\frac{R}{a}\right)
+2\sum_i\sum_{j>i} k_ik_j \log\left(\frac{a}{r_{ij}}\right)\right],\nonumber\\ 
\label{interaction}
\end{eqnarray}
with $r_{ij}$ the relative distance between the $i$th and $j$th defects.
From this we can see that the free energy of 
two interacting defects with opposite 
charges ($k_1+k_2=0$) does not depend on the system size $R$ 
so the logarithmic divergence when $R\to\infty$ disappears. 
Also it is obvious that the interaction energy from the second-term is a decreasing (increasing) function of $r_{12}$ when $k_1k_2>0$ ($k_1k_2<0$) implying that topological charges of same signs repel (attract) each other.
Also, the conservation of the total topological charge, equal to $q_t+6$, is equivalent to the constraint $\displaystyle\sum_{i=1}^N k_i=1$, i.e. 
the total winding number of the defects is equal to 1. Consequently the second term in (\ref{interaction}) is the only one responsible for the difference between the elastic free energies of different defect configurations. 

As mentioned before, the director field will be used to identify the charge of individual defects.
In the following we calculate such a field for the different lowest-order defects expected in our
system. Given the field $\theta({\bm r})=\theta(x,y)=\theta(\varphi)$, the curves whose tangent lines
have slopes $dy/dx=\tan{\theta}$ correspond to the one-parametric family of curves resulting
from the solution of the first-order equation
\begin{eqnarray}
\frac{dy}{dx}=\frac{r'\sin\varphi+r\cos\varphi}{r'\cos\varphi-r\sin\varphi}= \tan \theta=\tan (k\varphi+\theta_0).
\end{eqnarray}
where $r'=dr/d\varphi$. Solving for $r'$,
\begin{eqnarray}
r'=r\cot \left[(k-1)\varphi+\theta_0\right].
\end{eqnarray}
After integration we arrive at
\begin{eqnarray}
r(\varphi)=r_0\left|\sin\left[(k-1)\varphi+\theta_0\right]\right|^{1/(k-1)}.
        \label{uno}
\end{eqnarray}
The 6-atic field around defects with $k=\pm 1/6$ (selecting phases $\theta_0=0,\ \pm\pi/3$), 
i.e. with topological charges $q=\pm 1$, are shown in Figs. \ref{fig1}(a) and (b). We note the
characteristic pentagonal ($q=+1$) or heptagonal ($q=-1$) patterns about the defect, either convex ($q=+1$) or nonconvex ($q=-1$), depending on the sign of the charge.

Note that these director configurations refer to free topological defects. As we will promptly see, defects in the circular cavity are close to the cavity wall, but they can still be considered as free defects. By contrast, in the square cavity there are always defects which are pinned to the
corners of the square cavity; in this case it is necessary to redefine what we mean by winding number and topological charge. Because it is reasonable to expect that anchoring conditions at the walls are strong, a set of triangles will always orient with one of their three equivalent sides parallel to the walls, while their nearest neighbours (the complementary set) have one of their 
vertexes in contact with the wall and their heights perpendicular to it, in such a way that the two sets of
particles form a perfect monolayer. As a consequence, the director field is forced to be aligned 
perpendicular to the walls, and therefore rotates by $90^{\circ}$ at the corner. To this, we must add the rotation of the field from one wall to the other around the defect following a contour inside the cavity, which
accumulates an angle of $\pm 30^{\circ}$, see Figs. \ref{fig2}(a) and (b). 
The final result is that the defected structures at the corners can be interpreted as having
charges of $q=+1$ and $q=+2$. In fact, this result is equivalent to calculating the configuration 
of the director field around a free defect with an effective winding number of $\tilde{k}=\pm 1/3$, extracting the field 
structure from the specific quadrant shown in Figs. \ref{fig2}(a) and (b), and adding the constant wall rotation angle of $90^{\circ}$. The total rotation of the pinned defect is $90^{\circ}\pm 30^{\circ}$, 
giving angles of $60^{\circ}$ ($q=+1$) and $120^{\circ}$ ($q=+2$) and consequently the winding number of pinned defects are 
$\displaystyle{k=\frac{\pi/3}{2\pi}=\frac{1}{6}}$ ($q=+1$) and 
$\displaystyle{k=\frac{2\pi/3}{2\pi}=\frac{1}{3}}$ ($q=+2$). Free defects of charge $q=+2$ are
not expected to get stabilised in a 2D nematic due to their large elastic energy, but can be partially
screened when anchored to the corner of a square and, as seen in Section \ref{results}, help explain the global topological behaviour of the granular fluid in the square cavity. 

In the Appendix we generalize these topological rules to determine the charges of pinned defects and the number of possible defect configurations, including free ones, for any confining cavity shaped as a regular polygon with $n$ sides.

\begin{figure*}
\includegraphics[width=1.0\textwidth]{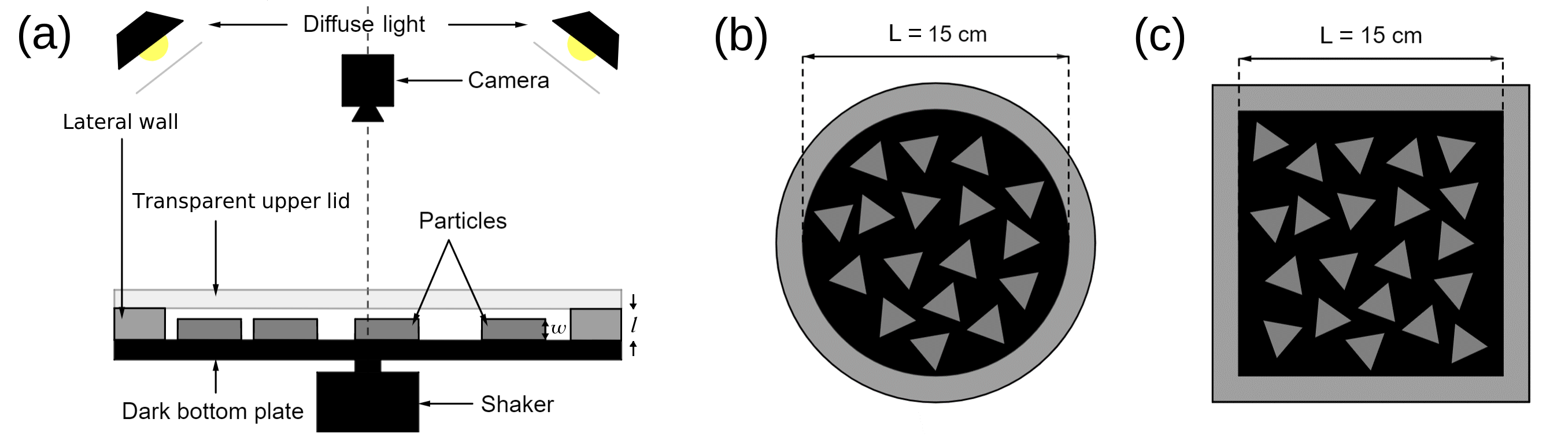}
\caption{\label{fig3} (a) Schematic of the experimental setup (lateral view). (b) Zenithal view of the circular cavity. (c) Zenithal view of the square cavity. }
\end{figure*}

\section{\label{experimental}Experimental setup and Monte Carlo simulation}

The experimental setup is similar to that used in previous work by our group \cite{Nosotros1,Nosotros2,Nosotros3}, and has been described 
elsewhere. In brief, particles made of nonmagnetic steel with the shape of an equilateral triangular prism are confined into
a horizontal cavity, Fig. \ref{fig3}, forming a quasimonolayer. A transparent upper lid made of methacrylate confines the
particles vertically into a space with a height $l$, with $w<l<2w$.
$w=1$ mm is the thickness of the triangular prisms. This allows particles to move vertically, while preventing two particles to stack one on top of the other. Their lateral size is $s=4$ mm, and $l$ must be smaller than the height of the particles, $s\sqrt{3}/2$, so that a particle cannot stand up on one of its edges. The cavities, either circular or square, have a size $L=15$ cm across, and are vibrated vertically
with frequency $\nu=100$ Hz, at an effective gravity such that the monolayer exhibits uniform density throughout the cavity. Particle configurations are recorded from above using a CCD camera
at a rate of 4 pictures per minute. 

Experiments typically run for 2-3 hours, and the system needs approximately 1 hour to reach the steady state. Particles are identified using a code written in {\tt Matlab}, which also extracts their position and orientation. The latter cannot be correctly identified
with the standard procedure of using the ellipse which best fits the set of pixels pertaining to the particle edges: a circle will always give the best fit for equilateral triangles and consequently no particle orientation can be extracted from the built-in {\tt Matlab} function. To solve this problem a new function was programmed which produced particle orientations as follows. The distance from the particle's barycenter to its edge, as a function of the angular variable in polar coordinates, can be calculated analytically for any given particle orientation angle (measured with respect to a fixed coordinate frame). Alternatively, this distance can be computed from the set of pixels defining the particle edges, using a function in {\tt Matlab}. By minimizing the sum of the squared differences between the analytical and experimental distances with respect to the particle orientation angle, an approximate value for the orientation angle can be obtained. This procedure yields highly accurate results for the particle's orientational field.
Using these orientations the set of order parameters are calculated (see Section \ref{results}), and color maps for the order parameters $Q_2$ and $Q_4$ are produced.

To better grasp the significance of the experimental results, we have also set up MC simulations on a similar system of model particles. By doing so, we initially expected to draw conclusions about the effects of dissipation and nonequilibrium phenomena on the rich behavior presented by the granular monolayer, and the possibility to demonstrate the similarities between the
two systems.
To generate systems with different packing fractions $\eta$, we first initialised the system with a lower packing fraction ($\eta\sim 0.6$), fixing the number of particles to $\sim 2000$ for both circular and square cavities (this is chosen so as to mimic the number of particles in the experiments).
Next we let the simulation run for some time, and
then repeatedly try to reduce the size of the container step by step, keeping all particles inside
the container. Then different systems with the desired packing fractions ($\eta\in[0.75, 0.85]$) can be prepared. Once a configuration at the desired density is obtained, we perform $2\times 10^5$ MC
steps (MCS), one MCS corresponding to an attempt to move all particles in the system sequentially.
Measurement of the relevant quantities explained in Section \ref{results} is made every 500 MCS,
totalling 400. Translational and rotational amplitudes (maximum displacement or angle a particle can be displaced or rotated) are monitored and changed to target an acceptance probability of 
$0.35$. Particle movements are set up so that on one third of the times
particles are only translated, on another third
particles are only rotated, and in the rest both movements are made simultaneously.

\section{\label{results}Results and discussion}

We present results for the circular and square cavities separately. In both cases a range of
packing fractions was explored, spanning most of the stability window of the 6-atic liquid crystal. 
In order to identify and characterize the different configurations in which particles can be arranged, we need some way to determine the orientational order of the system. First we define three orientational order parameters, $Q_n=\left|\langle e^{in\phi}\rangle\right|$, $n=2,4,6$, 
that probe two-, four- and sixfold symmetries, respectively. $\phi$ is the angle between the particle axis and a fixed reference frame, for instance the $x$ axis of the laboratory's reference frame. The angular brackets indicate that we take the average in a circular region of radius $4s$ centered on a particle. We measure this
quantity on each particle of the system. This expression of the order parameters follows from the second-rank tensor 
$\hat{Q}={\bm \langle 2{\bm e}\otimes {\bm e}-\mathbb{1}\rangle}$ of the nematic phase
(with ${\bm e}$ the unit vector pointing along the particle axis). The nematic director of uniaxial rods can be found by computing the eigenvector corresponding to the maximum eigenvalue of this tensor.
For particles with two (squares) or three (equilateral triangles) equivalent axes, generating the
4-atic or 6-atic phases, respectively, the orientational-order tensors are defined similarly, but they are of four
and six rank respectively \cite{Selinger}. The procedure to find the directors is exactly the same: the calculation of the
eigenvectors corresponding to the maximum eigenvalues of properly contracted tensors to obtain a 
second-rank
tensor \cite{Selinger}. In turn the maximum eigenvalue is the order parameter. Following this procedure one can find that
the maximum eigenvalue for the 6-atic symmetry is just $Q_6 = \left[\langle\cos{6\phi}\rangle^2 + 
\langle\sin{6\phi}\rangle^2\right]^{1/2}=\left|\langle e^{6i\phi}\rangle\right|$. Note that these order parameters are defined locally at each particle. The calculation of the order parameters is made for both the granular experiments and the MC simulations.

\subsubsection{\label{equilateral_circular}Circular cavity}

\begin{figure*}
\includegraphics[width=0.95\textwidth]{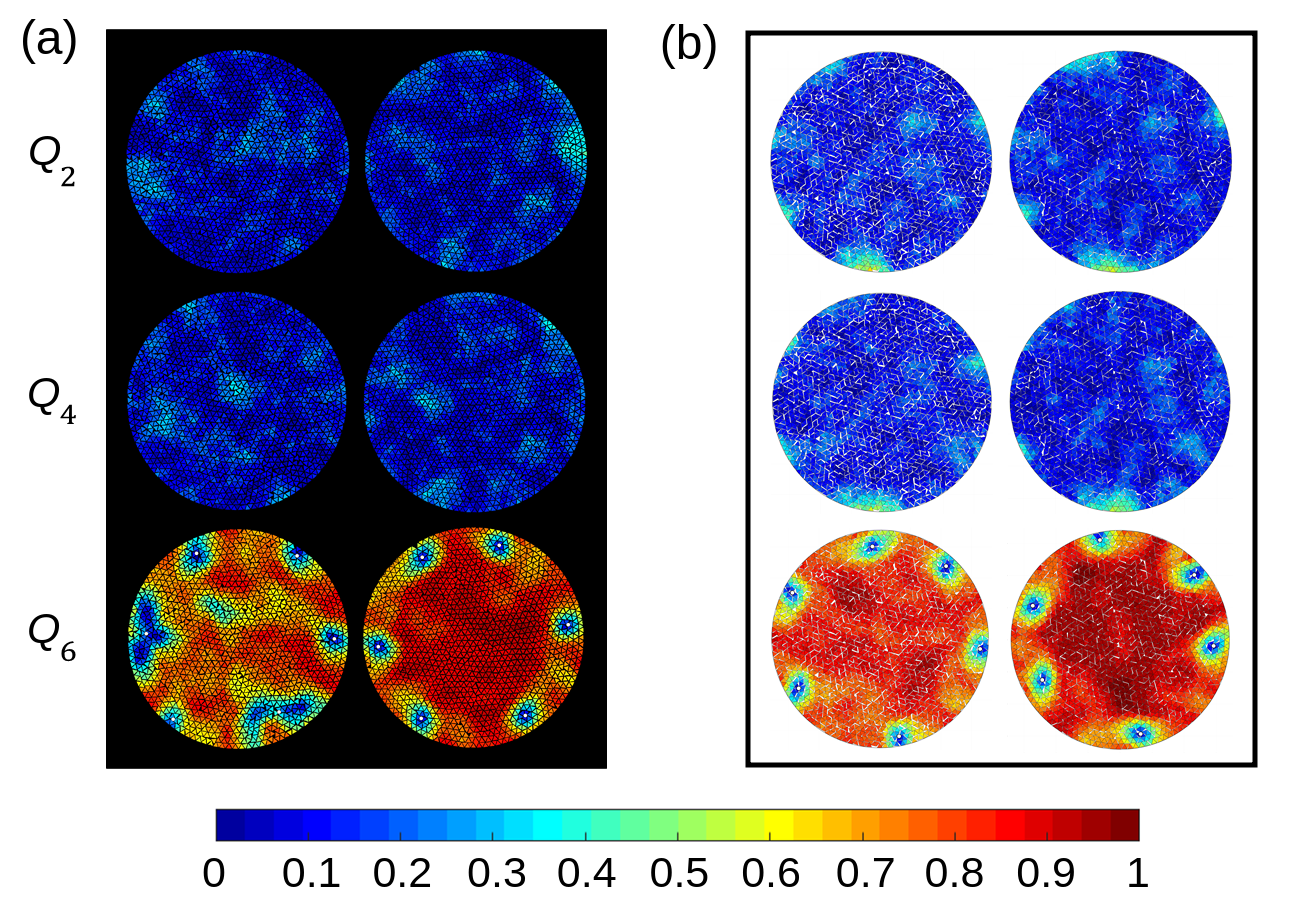}
\caption{\label{fig4} Order parameters. (a) Experiments, $\eta=0.78$ (left) and $0.85$ (right). (b) Simulations (same packing fractions). Value of the order parameters are indicated in the color bar.}
\end{figure*}

\begin{figure*}
\includegraphics[width=0.95\textwidth]{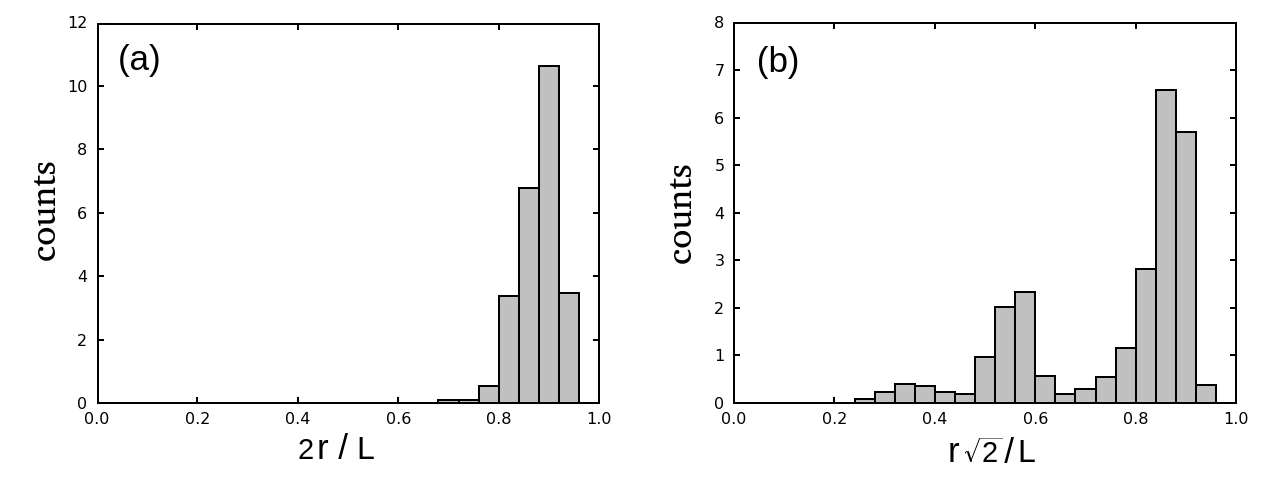}
\caption{\label{fig5} Histograms of defect positions. (a) Circular cavity, using the radius of the cavity $L/2$ as the unit of length. (b) Square cavity, using half of the cavity diagonal $L/\sqrt{2}$ as the unit of length.}
\end{figure*}

Fig.  \ref{fig4} shows representative colour maps of the three order parameters, from experiment and simulation, for
two different values of packing fraction. After a short equilibration time, a clear 6-atic phase is stabilised in the system in the steady state (low values for the $Q_2$ and $Q_4$ order parameters, and a high value of $Q_6)$. As expected, geometric frustation causes the system to develop six defects of charge $q=+1$, giving a total charge $q_{\rm t}=+6$. These defects, being
positive, repel each other through the orientational stiffness of the 6-atic fluid, and are located very close to
the inner wall of the cavity approximately at the vertexes of an hexagon 
inscribed in the cavity. This structure is subject to global rotation, as is typical in vibrated monolayers. Fig. \ref{fig5}(a) shows an example of a distribution of the radial positions of the defects, which is a sharply-peaked function located at $r\simeq 0.9L/2$.

As a complementary calculation, we have simulated a fluid of 2D hard triangles using MC
technique. Properties of the granular monolayers are reproduced in the MC simulations, demonstrating
that, despite their dissipative nature, the basic physics of the behaviour of the granular quasimonolayer
can be understood in terms of excluded-volume particle interactions, which create oriented fluids and
a stiff but deformable director field, much as in their thermal-equilibrium counterparts.
This is demonstrated in Fig. \ref{fig4}, which compares particle configurations in the granular and
thermal system for two different densities. 

\begin{figure}
\includegraphics[width=0.45\textwidth]{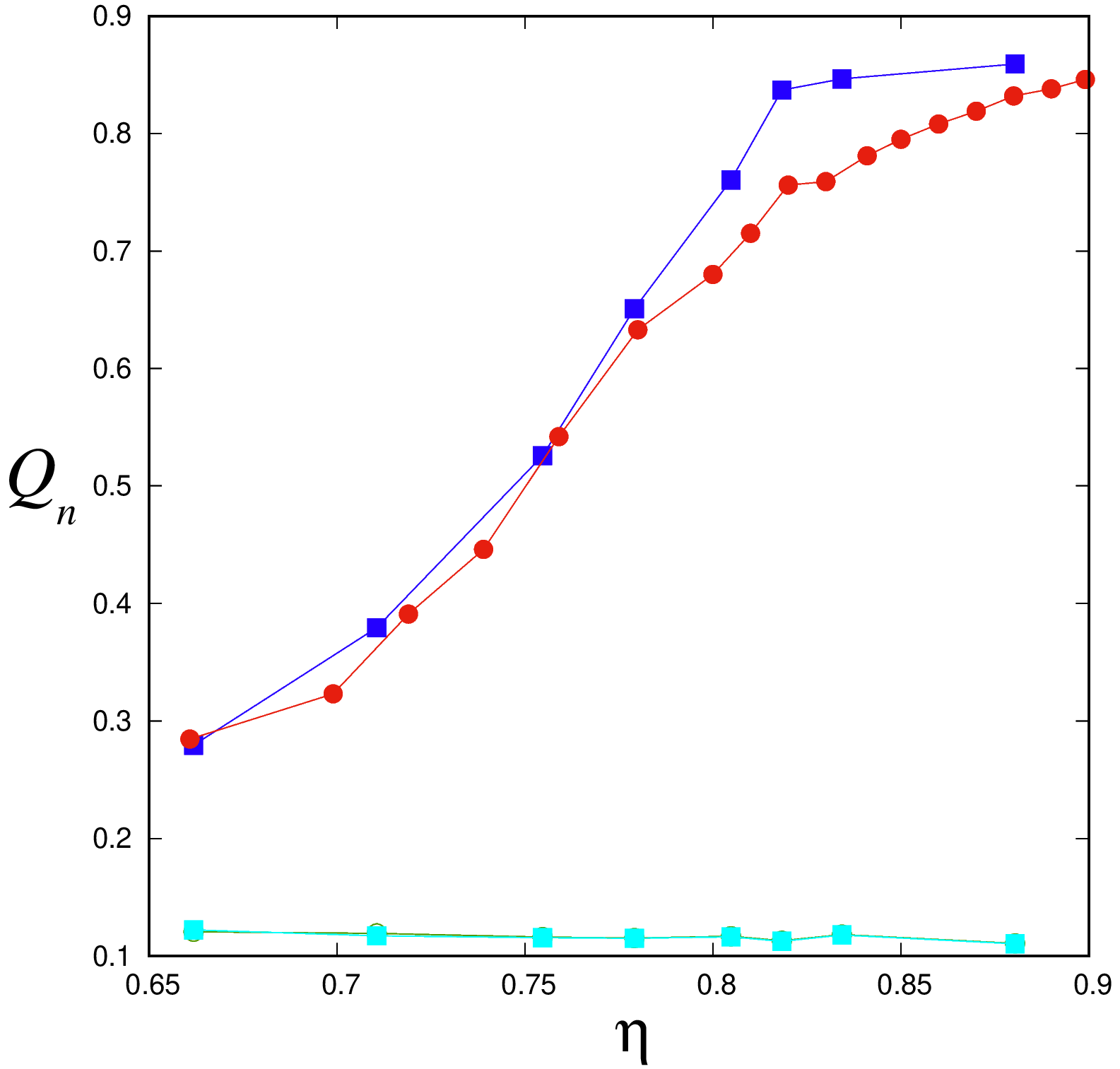}
\caption{\label{fig6} Order parameters $Q_n$, averaged over the whole cavity, as a function of the packing fraction $\eta$ in monolayers of 
equilateral triangular prisms confined in circular (circles) and square (squares) cavities. The 6-atic order 
parameter $Q_6$ is the one between 0.27 and 0.9 while the 
others $Q_n$ ($n=2,4$) are relatively small (close to 0.1). 
Note how the square confinement enhances the 6-atic 
ordering.}
\end{figure}

Fig. \ref{fig6} shows the behavior of 
the order parameters $Q_n$ with packing fraction in the granular monolayer. As expected 
from the six-fold symmetry of the particle, the order parameters 
$Q_2$ and $Q_4$ are vanishing small while $Q_6$, a direct measure 
of the 6-atic ordering, saturates at values $\sim 0.85$ when packing fraction is close to $0.9$
(note that the crystal phase expected at such high values of packing fraction does not appear in our experiments since the boundary conditions imposed by the circular confinement are incompatible 
with the triangular or hexagonal symmetries of the crystalline lattices). 
As usual $Q_6$ decreases as $\eta$ 
is lowered but with an interesting behavior: instead of having 
a relatively abrupt change close to the isotropic to 6-atic phase transition as in equilibrium systems, it decreases with a rather small slope to values clearly different from zero. This behavior is due to the dissipative nature of the system, which promotes the presence of a large proportion of clusters formed by 
hexagonal aggregates of triangles. In regions where  
clusters are present, the order parameter $Q_6$ 
is relatively high, resulting in a total averaged value (over the whole sample) of $Q_6$ appreciably different from zero.

\begin{figure*}
\includegraphics[width=0.75\textwidth]{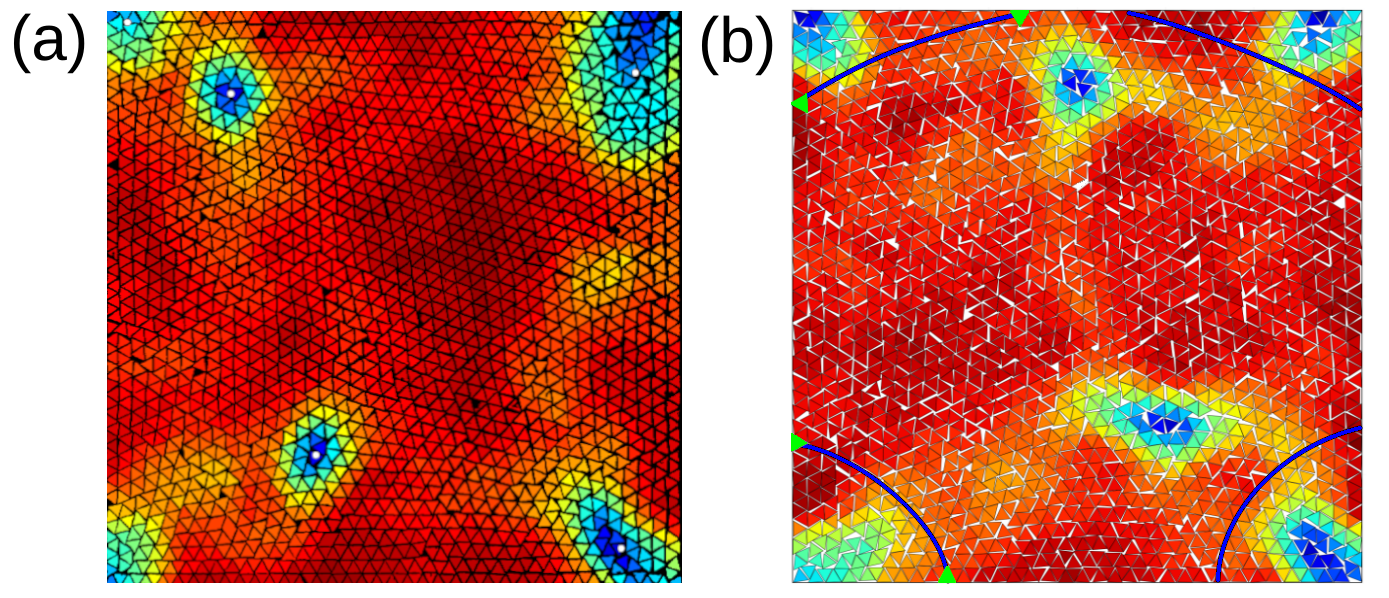}
\caption{\label{fig7} Comparison between configurations of triangular prisms in the square cavity as obtained from (a) granular monolayers (packing fraction $\eta=0.82$) and 
(b) MC simulations ($\eta=0.85$). In both cases two $+2$ charges are found in the square corners at the bottom, while two $+1$ charges are at the upper corners (cf. Fig. \ref{fig7}). Free defects with charges $+1$ and $-1$ are located in the bulk of the system (cf. Fig. \ref{fig1}). Approximate director lines around pinned defects of charges $+2$ and $+1$ are depicted.}
\end{figure*}

\subsubsection{\label{equilateral_square}Square cavity}

Steady-state configurations in the square cavity also comply with the rules of topology, but there
are some features that are not seen in the circular cavity. We start with 
Fig. \ref{fig6}, which shows the evolution of $Q_n$ with $\eta$. It 
is interesting to note that at high density the square confinement enhances 
the 6-atic ordering as compared with its circular counterpart. 
This is due to the strong anchoring effect of the square 
walls promoting a perfect orientation of 
triangles belonging to the first few monolayers in contact with the walls of the square. For the circular cavity the orientations 
of triangles at contact with the circular wall has 
a continuous variation due to the curvature of the container).

The distinct nature of the defects in the square cavity is shown in Fig. \ref{fig7}, which depicts
representative maps of $Q_6$. In this case six defects are excited, as in the circular cavity, but
not all defects have the same nature. We can distinguish two types of defects.
One type corresponds to \textit{pinned} defects, found in or close to the four corners of the cavity, which
present no motion.
These defects are always present as the corners are not compatible with a continuous director field with
6$-$atic symmetry. The other type of defects correspond to \textit{free} defects, which are excited in 
the bulk of the cavity and are not static but in fact very dynamic.
Free defects may appear in number as none, one, two or three (in Fig. \ref{fig7} two free
defects are visible, both in the granular quasimonolayer and in the MC simulations). 
The radial distribution of defects, shown in Fig. \ref{fig5}(b), 
reflects the existence of a strongly located population close to the corners
of the cavity (pinned defects), along with a group of defects loosely located
in the bulk of the cavity (free defects).

\begin{figure*}
\includegraphics[width=1\textwidth]{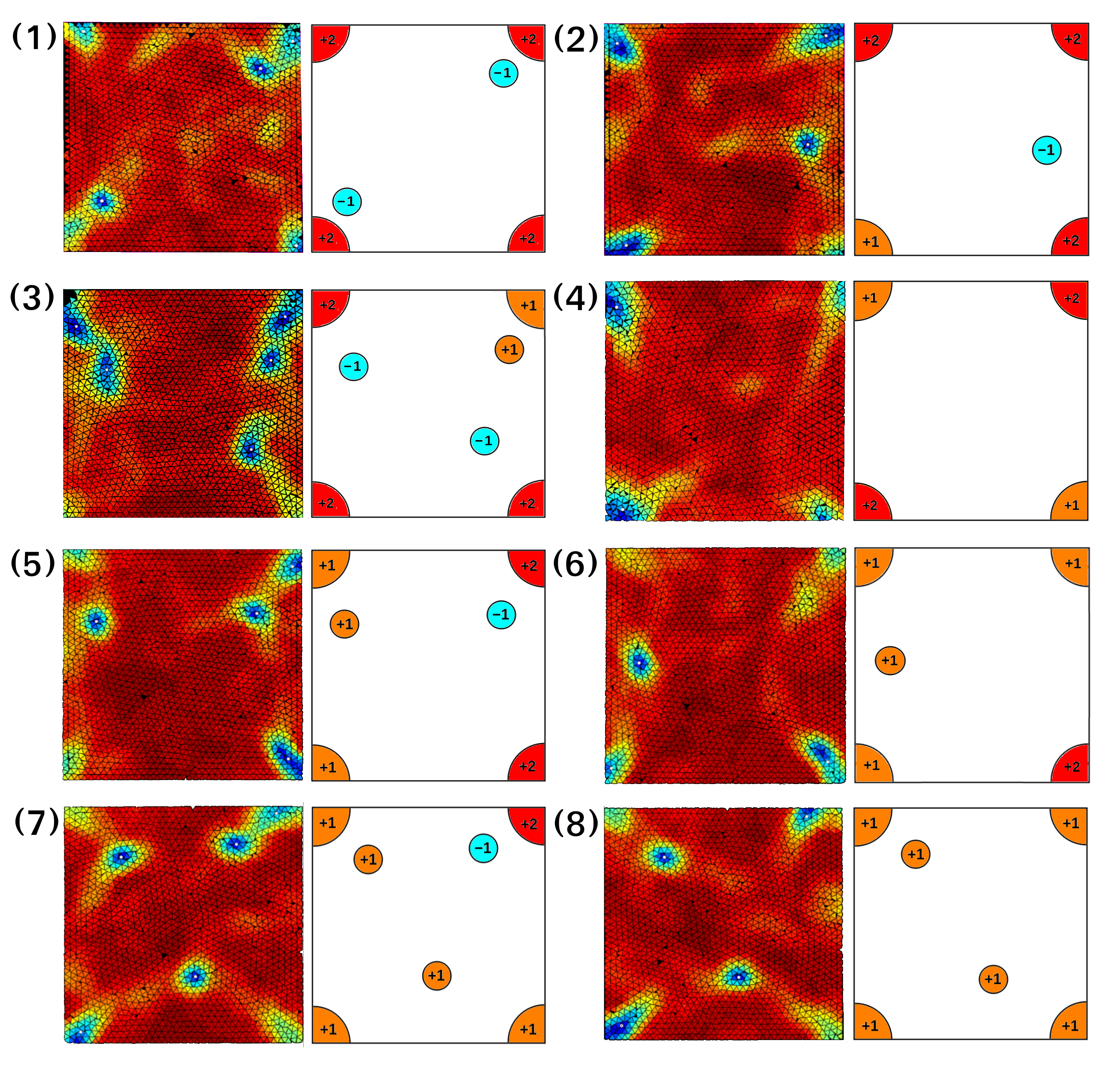}
\caption{\label{fig8} The different defect structures exhibited by the granular monolayer. The labels $1-8$ correspond to the configurations collected in Table \ref{table1}.}
\end{figure*}

One important feature 
of this system is that defects may have different charges, some of them even with different signs, and therefore the system may exhibit repelling and attractive  defect pairs. It will prove useful to designate different configurations by listing the set of their charges. In the following a given configuration of defects will be denoted by a list of all the defect charges,
starting with the pinned ones (indicated by a subscript `p') and followed by the free defects (subscript `f'). To shorten notation negative charges will be underlined.
Thus, the configurations shown in Fig. \ref{fig7} are both denoted by
$2_{\rm p}2_{\rm p}1_{\rm p}1_{\rm p}1_{\rm f}\underline{1}_{\rm f}$. Note that this notation does not distinguish between the cases where the two $+2$ charges are along the square diagonal or at the same side of the square.

Fig. \ref{fig8} presents a summary of all 
eight structures found in the granular system. Representative configurations of the granular system are shown, together with schematics depicting the charge and location of the defects in the images of the granular monolayers. 
Table \ref{table1} shows these combinations of pinned and free defects for the different configurations, along with their charge designations. 
The conservation of total topological charge requires the
charge of the individual defects to be correctly interpreted in all cases where varying numbers of
defects are present. An important point is that, from the analysis of Section \ref{theory}, all eight types of configurations can be explained in terms of four pinned defects at the corners of the cavity, with charges $+2$ or $+1$, and free defects in varying numbers with charge $\pm 1$. In this way all configurations satisfy the requirement of the total topological charge being $q_{\rm t}=+6$. As mentioned previously, the identification of defect charges can be made from the geometry of the director field in the neighborhood of the defects. Thus, as shown in Fig. \ref{fig7}, pinned defects of charge $+2$ are identified by director lines which are concave with respect to its corner, while pinned defects of charge $+1$ exhibit distinctive convex director lines. Free defects can similarly be identified by lines forming convex pentagonal shapes (charge $+1$) and nonconvex, heptagonal shapes (charge $-1$).

\begin{table*}
    \begin{tabular}{|c|c|c|c|c|}
    \hline
    \textbf{\# conf.} & \textbf{charges (pinned)} & \textbf{charges (free)} & ${\cal F}_{\rm el}/(\pi K)$ 
    & ${\cal F}_{\rm el}^{(\rm min)}/(\pi K)$ \\
    \hline
    1 & $2_{\rm p}$ $2_{\rm p}$ $2_{\rm p}$ $2_{\rm p}$ & $\underline{1}_{\rm f}$ $\underline{1}_{\rm f}$ &
    1.3548 & 1.3140\\
    2 & $2_{\rm p}$ $2_{\rm p}$ $2_{\rm p}$ $1_{\rm p}$ & $\underline{1}_{\rm f}$ & 1.3268 & 1.2745\\
    3 & $2_{\rm p}$ $2_{\rm p}$ $2_{\rm p}$ $1_{\rm p}$ & $\underline{1}_{\rm f}$ $\underline{1}_{\rm f}$ $1_{\rm f}$ & 1.3280 & 1.2564\\
    4 & $2_{\rm p}$ $2_{\rm p}$ $1_{\rm p}$ $1_{\rm p}$ & $0_{\rm f}$ & 1.2365 & 1.2365\\
    5 & $2_{\rm p}$ $2_{\rm p}$ $1_{\rm p}$ $1_{\rm p}$ & $1_{\rm f}$ $\underline{1}_{\rm f}$&1.2825 & 1.2300\\
    $5'$ & $2_{\rm p}$ $2_{\rm p}$ $1_{\rm p}$ $1_{\rm p}$ &
    $1_{\rm f}$ $\underline{1}_{\rm f}$ & X & 1.2197\\
    6 & $2_{\rm p}$ $1_{\rm p}$ $1_{\rm p}$ $1_{\rm p}$ & $1_{\rm f}$&1.2208 & 1.2048\\
    7 & $2_{\rm p}$ $1_{\rm p}$ $1_{\rm p}$ $1_{\rm p}$ & $1_{\rm f}$ $1_{\rm f}$ $\underline{1}_{\rm f}$&1.2708 & 1.1848\\
    8 & $1_{\rm p}$ $1_{\rm p}$ $1_{\rm p}$ $1_{\rm p}$ & $1_{\rm f}$ $1_{\rm f}$&1.2209 & 1.1676\\
    \hline
    \hline
    \end{tabular}
    \caption{\label{table1} Different defect structures containing four pinned defects and a number
    of free defects observed in the granular monolayer. The configuration number corresponds to the number assigned in Fig. \ref{fig8}. Pinned charges are indicated with a `p'
    subscript, whereas an `f' subscript refers to free charges. Negative charges are underlined.
    Note that configuration \# 4 displays no free charges.
    The 4th column is the elastic free energy in units of $\pi K$ from Eqn. (\ref{interaction}) corresponding to configurations of Fig. \ref{fig8} while 
    the 5th collect the minima of the elastic free-energy from the same equation where the distances between charges 
    were taken to be greater or equal to $2a$. The configurations corresponding to these minima are shown in Fig. \ref{fig9}. Configuration $5'$ is not found in the experiments, and refers to a charge distribution similar to that of configuration 5 in terms of location of pinned and free charges, but with pinned charges of $+2$ located opposite along the diagonal of the cavity (see Fig. \ref{fig8}). According to elastic theory the free energy of
    configuration $5'$ is lower than that of configuration 5.}

\end{table*}

To see if we can rationalise this
phenomenology in terms of the effective free energies implied by elastic theory, we consider the elastic interaction energy (\ref{interaction}) and a simplified model for defect cores. Assuming defect cores to behave as hard particles of diameter $2a$, we can search for the minimum-energy position of free charges in each configuration, keeping the pinned charges at the corner positions. We explore the multiparametric space of the coordinates $(x_i,y_i)$, $i=1,\cdots,n$, with $n$ the number of free charges in each configuration, both using a discrete grid and alternatively using a MC sampling scheme. Note that we assume pairwise interactions, and do not try to model specific interactions between defects at short distances, which would require the minimisation of a full elastic model taking defects as singularities of the local orientation field.
The results from the simplified elastic model does not fully grasp the fine details of the granular configurations, but help understand the defect configurations. 

Table \ref{table1} shows values of the elastic free energy of configurations of charges shown in Fig. \ref{fig8}. 
Also it is collected in the same table the minimum-energy charge configuration,
obtained by minimising the elastic free energy of the corresponding $N$-defect systems using Eqn. (\ref{interaction}). 
The parameters have values $R=2 L$ (with $L$ the edge-length of 
the square cavity) and $a=0.1 L$. Pinned defects are taken to be static and have 
been located at the vertexes of the cavity. The ordering of the free-energies (from lower to higher) corresponding to defect configurations from Fig. \ref{fig8} produce the sequence $6\to 8\to 4\to 7\to 5\to 2\to 3\to 1$. However the ordering of the minimum free-energies 
produce the following one: $8\to 7\to 6 \to 5 \to 4 \to 3 \to 2\to 1$, implying that the free-energy is higher when the number of $+2$ charges is larger.
 
\begin{figure*}
\includegraphics[width=1.0\textwidth]{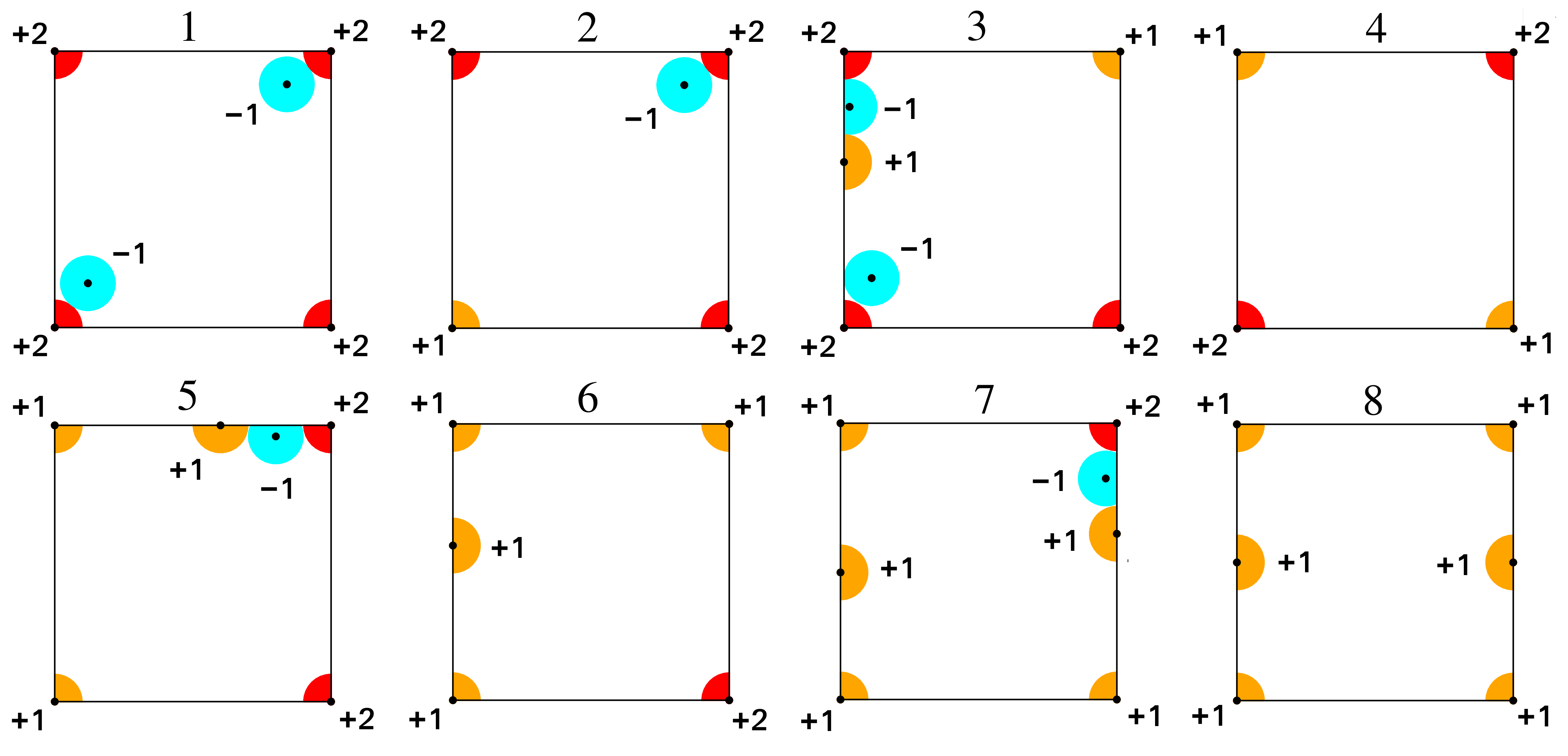}
\caption{Equilibrium distribution of charges obtained from the minimization of Eqn. (\ref{interaction}) selecting $R=2L$ and $a=0.1L$ with pinned and free charges being those of 
configurations shown in Fig. \ref{fig8}. The pinned charges are always located at the vertexes of the square and we selected $2a$ 
as the minimum distance between any two charges. 
 }
\label{fig9}
\end{figure*}
Fig. \ref{fig9} presents a schematic diagram with the location of charges in the minimum energy configurations.
In general, when a negative free charge is 
present, it sticks close together to one of the pinned $+2$ charges. 
If, in addition to the negative free charge, there is a positive charge, 
the two stick together. When all pinned charges have a charge of $+2$, the two  
negative free charges stay in contact with two pinned charges 
located at one of the square diagonals, with their centers of mass positioned close to the diagonal. Finally when all the charges have 
positive $+1$ charge, free charges are located midway between 
a pair of pinned charges along one of the sides of the square, and one in front to the other. Therefore, in most cases defects are located close or at the sides of the square.   

As regards the similarities between experiment 
and theory as far as the position of free defects is concerned, a simple comparison between Figs. \ref{fig8} and \ref{fig9} shows that configurations 1 and 6 are very similar. As a general rule, experiment and theory both exhibit configurations where negative free charges are very close to a pinned defect of charge $+2$. However, the predicted tendency of defects to be close to the boundaries according to elastic theory does not hold in the case of experiments (see Fig. \ref{fig8}), where free charges tend to be in the bulk of the cavity, far from the boundaries.
This difference can be explained by the presence of layers 
of triangles close to the walls featuring strong crystalline ordering (due to the strong anchoring conditions at the walls), which reflects 
in an effective repulsive interaction between these layers and the defects, a fact not taken into account by the model. 
Also, when a negative free charge is present, the minimum free-energy configuration predicted by the elastic model is that corresponding to the coalescence of the negative charge with one $+2$ pinned charge, 
except when certain constraint for a minimum distance between charges is imposed, something that we have done by assuming $r_{ij}\geq 2a$. 
In experiments negative charges are close to the $+2$ charges most of the time, but can travel to relatively distant positions (see configuration 2 in Fig. \ref{fig8}).

Note that the configurations of the granular monolayers shown in Fig. \ref{fig8} do not necessarily correspond
to minimum energy configurations since free defects are very dynamic and, as discussed below, undergone fusion and disintegration events, which make the system to transit from one to another of the typical configurations shown in Table \ref{table1}. Also, the elastic model
only represents an approximation to the real physics, and minimum-energy 
configurations for a system with a particular number and distribution of charges
may not necessarily be related to the actual typical or average configuration with
the same charges. In particular, the pair interaction model implicit in
Eqn. (\ref{interaction}) may not be correct, particularly at short distances, as interactions between defects are
mediated by the rigidity of the 6-atic underlying phase, and collective interactions
are expected. 

Not only do different defect configurations stabilise in the system, but we observe a fast dynamics between different
types of configurations, meaning that these defects can undergo processes of fusion and splitting, and that the
system can visit multiple configurations (see Table \ref{table1}). Several such elemental events
can be seen in a given run. For example, 
\begin{eqnarray}
    &&\hbox{7} \longleftrightarrow \hbox{8}\hspace{0.4cm}\hbox{and}\hspace{0.4cm}\hbox{5} \longleftrightarrow \hbox{6}
    \hspace{0.4cm}\hbox{($+2_{\rm p}-1_{\rm f}\longleftrightarrow +1_{\rm p}$)}\nonumber\\
    &&\hbox{5} \longleftrightarrow \hbox{7}\hspace{0.4cm}\hbox{($+2_{\rm p}\longleftrightarrow +1_{\rm p}+1_{\rm f}$)}\nonumber\\
    &&\hbox{6} \longleftrightarrow \hbox{7}\hspace{0.4cm}\hbox{($0_{\rm f}\longleftrightarrow +1_{\rm f}-1_{\rm f}$)}\nonumber\\
    &&\cdots
\end{eqnarray}

\begin{figure*}
\includegraphics[width=6.in]{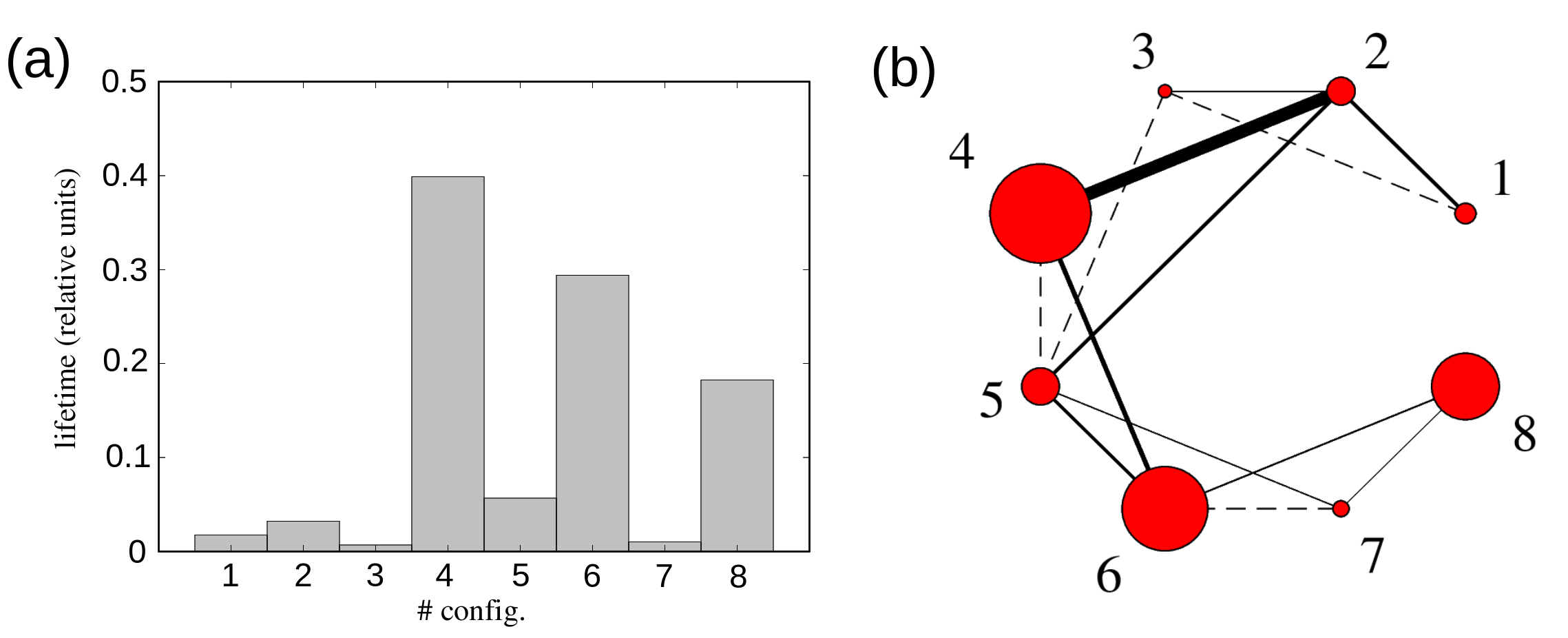}
\caption{(a) Histogram of 
the lifetime for each configuration (in relative units). (b) Network  diagram representing different configurations (nodes) of topological charges in the square cavity, along with transitions (links) between them.
The area of nodes is proportional to the lifetime of the configuration, while the width of the links indicates the relative number of (reversible) reactions between a given connected pair of nodes. Dashed links indicate possible transitions not found in our experiments.}
\label{fig10}
\end{figure*}

The histogram of the lifetime for different 
configurations is shown in  Fig. \ref{fig10}(a). This is a measure of the degree the stability of each configuration. Clearly configurations 4, 6 and 8 are the most stable. The remaining configurations have short lifetimes, which may be compatible with transition states. 
The network diagram shown in Fig. \ref{fig10}(b) contains elemental `reactions' between different configurations found in the experiments. The nodes represent the configurations, with their area being proportional to the lifetime of the specific configuration. The links between the nodes represent the (reversible) transitions between them, and their width is made proportional to the number of transitions between the corresponding states found in the experiments. Dashed links joining states 1 and 3, and 3 and 5, indicate that, even though these transitions are possible in principle, they have not been found in our experiments (probably due to their finite duration). A possible explanation is the short lifetime of configuration 3. The diagram of panel (b) shows that configuration 4 (without free charges) is the most stable in terms of the lifetime followed by configurations 6 and 8. Note that configuration 4 has no free charges (see Fig. \ref{fig8}), while 6 and 8 do not exhibit the presence of negative free charges, a result which agrees with elastic theory in that the presence of negative charges is energetically penalized.

Table \ref{table2} contains all possible elemental `reactions' in terms of starting and ending configuration. Most of these
events can be seen in the experiments as shown in Fig. \ref{fig10} (b). The persistence of a given configuration has also be quantified. 
The relatively fast 
dynamics between defect configurations imply that the charges easily react between them. We have even
observed events where charges of the same sign ($+1_{\rm f}$ and $+1_{\rm p}$) fuse together to give a pinned defect of higher
charge ($+2_{\rm p}$):
\begin{eqnarray}
    &&\hbox{8} \longleftrightarrow \hbox{6}\hspace{0.4cm}\hbox{($+1_{\rm p}+1_{\rm f}\longleftrightarrow +2_{\rm p}$)}
\end{eqnarray}
This process might be possible due to the long-range repulsive interactions due to the rest of charges, which push
the free $+1_{\rm f}$ charge against the pinned $+1_{\rm p}$ charge. It would mean that the system may exist in a number of
equally metastable configurations with respect to some effective free energy. 
Also, we observe some correlation between the positions of the $+2_{\rm p}$ pinned charges. In configuration \#4 no free charges are
excited, and $+2_{\rm p}$ defects are located at opposite corners of the cavity, due to their high repulsion. Two $+2_{\rm p}$ defects
may occupy adjacent corners, but only when a $-1_{\rm f}$ free defect screens the interaction (see configuration \# 5).

\begin{table*}
    \begin{tabular}{|c|c|c|c|c|c|c|c|c|}
    \hline
    \textbf{\# conf.} & 1 & 2 & 3 & 4 & 5 & 6 & 7 & 8\\
    \hline
    1 & X & $2_{\rm p}\underline{1}_{\rm f}\rightarrow 1_{\rm p}$ & ${\color{gray} 2_{\rm p}\rightarrow 1_{\rm p}1_{\rm f}}$ & X& X& X& X&X\\
    \hline
    2 &$1_{\rm p}\rightarrow 2_{\rm p}\underline{1}_{\rm f}$ & X & $0_{\rm f}\rightarrow 1_{\rm f}\underline{1}_{\rm f}$ & $2_{\rm p}\underline{1}_{\rm f}\rightarrow 1_{\rm p}$&
    $2_{\rm p}\rightarrow 1_{\rm p}1_{\rm f}$&X&X&X\\
    \hline
    3 & ${\color{gray} 1_{\rm p}1_{\rm f}\rightarrow 2_{\rm p}}$ & $1_{\rm f}\underline{1}_{\rm f}\rightarrow 0_{\rm f}$& X & X & ${\color{gray} 2_{\rm p}\underline{1}_{\rm f}\rightarrow 1_{\rm p}}$ & X & X & X\\
    \hline
    4 & X &$1_{\rm p}\rightarrow 2_{\rm p}\underline{1}_{\rm f} $ &X & X & {\color{gray}$0_{\rm f}\rightarrow 1_{\rm f}\underline{1}_{\rm f}$} & $2_{\rm p}\rightarrow 1_{\rm p}1_{\rm f}$ & X & X\\
    \hline
    5 & X & $1_{\rm p}1_{\rm f} \rightarrow 2_{\rm p}$& ${\color{gray} 1_{\rm p} \rightarrow 2_{\rm p}\underline{1}_{\rm f}}$& {\color{gray}$1_{\rm f} \underline{1}_{\rm f} \rightarrow 0_{\rm f}$}& X & 
    $2_{\rm p}\underline{1}_{\rm f}\rightarrow 1_{\rm p}$ & $2_{\rm p}\rightarrow 1_{\rm p}1_{\rm f}$ & X\\
    \hline
    6 &X & X & X &$1_{\rm p}1_{\rm f}\rightarrow 2_{\rm p}$ & $1_{\rm p}\rightarrow 2_{\rm p}\underline{1}_{\rm f}$& X & {\color{gray} $0_{\rm f}\rightarrow 1_{\rm f}\underline{1}_{\rm f}$} & $2_{\rm p}\rightarrow 1_{\rm p}1_{\rm f}$\\
    \hline
    7 &X &X &X &X & $1_{\rm p}1_{\rm f} \rightarrow 2_{\rm p}$& {\color{gray}$1_{\rm f}\underline{1}_{\rm f}\rightarrow 0_{\rm f} $} & X &$2_{\rm p}\underline{1}_{\rm f}\rightarrow 1_{\rm p}$\\
    \hline
    8 &X &X &X &X & X& $1_{\rm p}1_{\rm f}\rightarrow 2_{\rm p}$&$1_{\rm p}\rightarrow 2_{\rm p}\underline{1}_{\rm f}$ & X\\
    \hline
    \end{tabular}
    \caption{\label{table2} Matrix of elemental `reactions' between the configurations. The symbol `X' indicates that there is no pathway between the corresponding configurations. Labels in gray correspond to possible reactions that have not been detected in our experiments.}
\end{table*}

The MC simulations exhibit a defect configuration which is similar to one of the configurations found in the granular monolayer. Fig. \ref{fig7} shows a comparison between the two, with the same defect configuration $2_p2_p1_f\underline{1}_f$. However, defect `dynamics' is very slow in the simulations, and we have not been able to detect large defect difusion and consequently fusion or disintegration events. This prevents from drawing any conclusion as to the effect of dissipation in the defect dynamics described. 

\section{\label{conclusions}Conclusions}

Topology predicts that the total charge in a 6-atic phase confined within circular and square cavities should be the same, $q_{\rm t}=+6$, since a circle and a square share the same topological properties. Vibrated granular monolayers indeed conform to this prediction. However, the distribution of defects —specifically their number and charge— differs significantly between the two geometries. This difference arises due to the presence of corners in the square cavity, which anchor two distinct types of defects. We have demonstrated that these static defects can have charges $q=+1_{\rm p}$ or $q=+2_{\rm p}$, and multiple charge combinations are possible. However, only one combination, 
$q=\{+2_{\rm p}, +2_{\rm p}, +1_{\rm p}, +1_{\rm p}\}$, is free from defects in the bulk and in fact the most frequently seen configuration in the experiments, probably being the most stable.

When the sum of the charges of the pinned defects is less than $+6$, free defects are excited in the bulk of the cavity to restore charge conservation. These free defects can have charges $q=\pm 1_{\rm f}$, meaning both positive- and negative-charge defects can be excited in the system. The coexistence of positive and negative charges in the system leads to fast dynamics, with rapid transitions between different configurations. The elastic model explains this by showing that configurations with different charge numbers have nearly equivalent elastic free energies. This does not imply that the free-energy barriers are particularly high or low; rather, experimental findings suggest that the barriers are relatively low, facilitating the fusion and disintegration of defects within the system.

The softness of this granular system arises from the presence of 6-atic orientational order within an otherwise fluid monolayer, as well as from the pinning effects of the sharp, right-angled corners of the cavity. These corners anchor two distinct types of pinned defects with charges $+1$ and $+2$. This is in contrast with the circular cavity, where only $+1$ free charge defects are excited. The unique geometry of the square cavity, combined with the six-fold orientational order in the fluid, provides an ideal setting for studying defect dynamics. While nonequilibrium processes, such as friction and dissipation, likely contribute to the fast defect dynamics observed, we have been unable to fully analyze their impact through comparison with equilibrium Monte Carlo (MC) simulations.

Our results may open up a new avenue in the study of defect configurations in confined liquid-crystalline systems and the effect of the geometry of the container, such as general $n$-sided regular or nonregular polygonal cavities. Some ideas related to defect configurations in systems of equilateral triangular particles confined in $n$-sided regular polygonal cavities are provided in the Appendix. 

\begin{acknowledgments}
Financial support from Grants PID2023-148633NB-I00/AEI 
and PID2021-126307NB-C21/MICIU/AEI/10.13039/501100011033/FEDER,UE is acknowledged. Ariel Díaz de Armas is gratefully acknowledged for his contributions to the software used in this work.
\end{acknowledgments}

\appendix
\section{Equilateral triangular prisms confined in polygonal cavities}
\label{appendix}

\begin{figure*}
\includegraphics[width=1.0\textwidth]{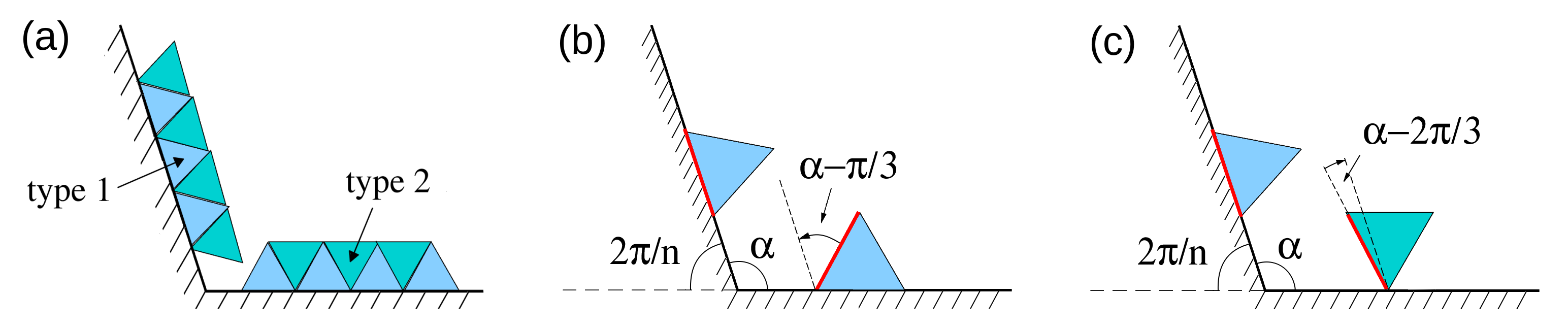}
\caption{(a) Schematic of two monolayers of equilateral 
triangles in contact with two consecutive walls of a polygonal cavity. Triangles with different orientations with respect to the walls have been indicated as `type 1' and `type 2'. (b) and (c) Rotation of a triangle close to a vertex of a $n$-sided confining polygon. In (b) the type is invariant under rotation, while in (c) the type changes.} 
\label{fig11}
\end{figure*}
In this paper we have investigated the ordering properties of a fluidized monolayer of equilateral triangular prisms confined within two types of cavities: a square cavity (with $n=4$ sides) and a circular cavity, which can be considered the limiting case of a regular polygon with $n\to\infty$ sides. Although both cavities share the same topological characteristics, namely the same total topological charge, we have found that the nature and spatial distribution of topological defects differs significantly between them.

This observation naturally raises the question: how does the distribution of defects evolve as the number of sides $n$ increases, and is there a critical value of $n$ beyond which the system exhibits the same behavior as in the circular cavity, characterized by six free defects each with topological charge $+1$? To address this, the present Appendix explores the general case of a regular polygonal cavity with arbitrary $n$, aiming to understand the crossover between polygonal and circular confinement.

We begin by demonstrating that defects are expected to form at the vertices of a polygonal cavity, and that only two types of defects can arise, with topological charges $q=+1$ and $+2$. To understand this, consider two adjacent linear boundaries that meet at a vertex of an $n$-sided polygon. Due to packing considerations, a well-ordered monolayer tends to form along each boundary, consisting of triangles arranged in an alternating pattern. In this configuration, particles adopt two distinct orientations: one where a triangle aligns with a side in contact with the boundary (`type 1'), and another where a vertex of the triangle touches the boundary (`type 2'), see Fig. \ref{fig11}(a).
Note that the the interior angle at the vertex of a regular $n$-sided polygon is $\alpha=\pi-2\pi/n$.

As one moves from the horizontal boundary to the oblique one following a director line joining triangles of the same type, the particle axis rotates $\displaystyle{\alpha-\pi/3=2\pi/3-2\pi/n}$ counterclockwise. See Fig. \ref{fig7}(b), where relevant director lines joining triangles at the boundaries for the pinned defects are depicted. For triangles of different types the rotation angle is $\displaystyle{\alpha-2\pi/3=\pi/3-2\pi/n}$, clockwise for $n<7$ and counterclockwise for $n\geq 7$ (see Fig. \ref{fig11} (b) and (c)). Both rotations are interior, i.e. they take place close but not at the polygonal vertex. Right at the vertex the rotation angle is $2\pi/n$, so that the total (accumulated) rotation angle is $2\pi/3$ and $\pi/3$, respectively. The associated winding numbers of defects pinned at the vertices of the $n$-sided polygon are then
\begin{eqnarray}
&&\displaystyle{k=\frac{(2\pi/3)}{2\pi}=\frac{1}{3}}\hspace{0.4cm}(\hbox{charge }q=pk=+2),\nonumber\\
&&\displaystyle{k=\frac{(\pi/3)}{2\pi}=\frac{1}{6}}\hspace{0.4cm}(\hbox{charge }q=pk=+1) 
\end{eqnarray}
where we set $p=6$ as appropriate for a $6$-atic symmetry. %{\color{red} The interior rotation with angle $\alpha-2\pi/3$, 
%compatible with the interchange $1\leftrightarrow 2$ of each triangular type belonging to clusters 
%of perfect monolayers at different walls, 
%has as a consequence that these clusters rotate as a whole from the horizontal to the oblique side. Of course the interior 
%rotation, with angle $\alpha-\pi/3$, that keeps unchanged the types of triangles of clusters at different walls share the 
%same property: a synchronic interior rotation of all triangles belonging to these clusters.}
It is important to note that for $n=3$ (triangular cavity) and 
$n=6$ (hexagonal cavity) triangles of the same (different) types rotate 
$0$ ($-\pi/3$) and $\pi/3$ ($0$) respectively, implying that 
no defects are present in these cavities. For other polygonal confinements defects of topological charges $+1$ and $+2$ will be present. 

\begin{table}
\begin{tabular}{|c|c|c|c|c|c|c|c|c|}
\hline
Conf. &  $\ 1\ $  &  $\ 2\ $  &  $\ 3\ $  &  $\ 4\ $  &  $\ 5\ $  &  $\ 6\ $  &  $\ 7\ $  &  $\ 8\ $ \\
\hline
$m_1$ & 5 & 5 & 4 & 4 & 3 & 3 & 2 & 1\\
\hline
$m_2$ & 0 & 0 & 1 & 1 & 2 & 2 & 3 & 4\\
\hline
$j_1$ & 1 & 2 & 0 & 1 & 0 & 1 & 0 & 0\\
\hline
$j_{-1}$ & 0 & 1 & 0 & 1 & 1 & 2 & 2 & 3\\
\hline
\end{tabular}
\caption{Defect configurations (assuming no more than three free charges) inside a pentagonal ($n=5$) cavity consisting 
of $m_1$ pinned defects of charge $+1$, $m_2$ pinned 
defects of charge $+2$, $j_1$ free defects of charge $+1$, and 
$j_{-1}$ free defects of charge $-1$.}
\label{table3}
\end{table}

Let us now consider the contribution of free charges. In an $n$-sided polygonal cavity, the sum of $m_1$ pinned defects with charge $+1$ and $m_2$ pinned defects with charge $+2$ must satisfy the condition
$m_1+m_2=n$. In the presence of $j_1$ and $j_{-1}$ free charges with values $+1$ and $-1$, respectively, the overall topological charge conservation leads to
\begin{eqnarray} 
m_1 + 2m_2 + j_1 - j_{-1} = 6 \quad \Rightarrow \quad 2n + j_1 - m_1 - j_{-1} = 6.\nonumber\\ 
\end{eqnarray}
In the specific case of a square cavity ($n = 4$), this condition simplifies to
\begin{eqnarray}
    2+j_1=m_1+j_{-1}.
\end{eqnarray}
By examining all possible values of $m_1$ in the range $0$ to $4$, we can identify the corresponding pairs $(j_1, j_{-1})$ that satisfy the charge balance equation. These combinations yield the distinct configurations of pinned and free charges which have already been discussed in detail in Sec. \ref{equilateral_square}.

Our analysis is restricted to configurations involving no more than three free charges, as configurations with higher numbers are highly unlikely in our system. This limitation stems from the fact that the cavity size is comparable to the extrapolation length of the liquid-crystalline 6-atic phase. Indeed, such excess charges are not observed in experiments. However, larger cavities could in principle support a greater number of free charges.

This approach can be extended beyond the square cavity. For instance, in the case of a pentagonal cavity ($n=5$), the charge conservation equation becomes
\begin{eqnarray}
4+j_1=m_1+j_{-1}.
\end{eqnarray} 
Applying the same reasoning and again restricting the number of free charges to at most three, we can enumerate all compatible configurations of pinned and free charges. These are summarized in Table \ref{table3}. Among these, configurations 1, 3 and 5  are expected to have the longest lifetimes as they contain either no free charges or only one, which minimizes the free-energy cost, as discussed previously. 

\begin{table}
\begin{tabular}{|c|c|c|c|c|c|c|c|}
\hline
$\ n\ $ & $\ 4\ $ & $\ 5\ $ & $\ 7\ $ & $\ 8\ $ & $\ 9\ $ & $\ 10\ $ & $\ \infty\ $\\
\hline
$C$ & 8 & 8 & 4 & 2 & 1 & 0 & 1\\
\hline
${\cal N}_{\rm min}$ & 4 & 5 & 8 & 10 & 12 & X & 6\\
\hline
${\cal N}_{\rm max}$ & 7 & 8 & 10 & 11 & 12  & X & 6\\
\hline
\end{tabular}
\caption{Total number of defect configurations $C$ with 
no more than then free charges, 
and minimum ${\cal N}_{\rm min}$ and maximum ${\cal N}_{\rm max}$ number 
of defects as a function of the number of sides of the polygon, $n$. 
The case of a circular cavity, $n\to\infty$, is also included.}
\label{table4}
\end{table}

\begin{figure*}
\includegraphics[width=0.95\textwidth]{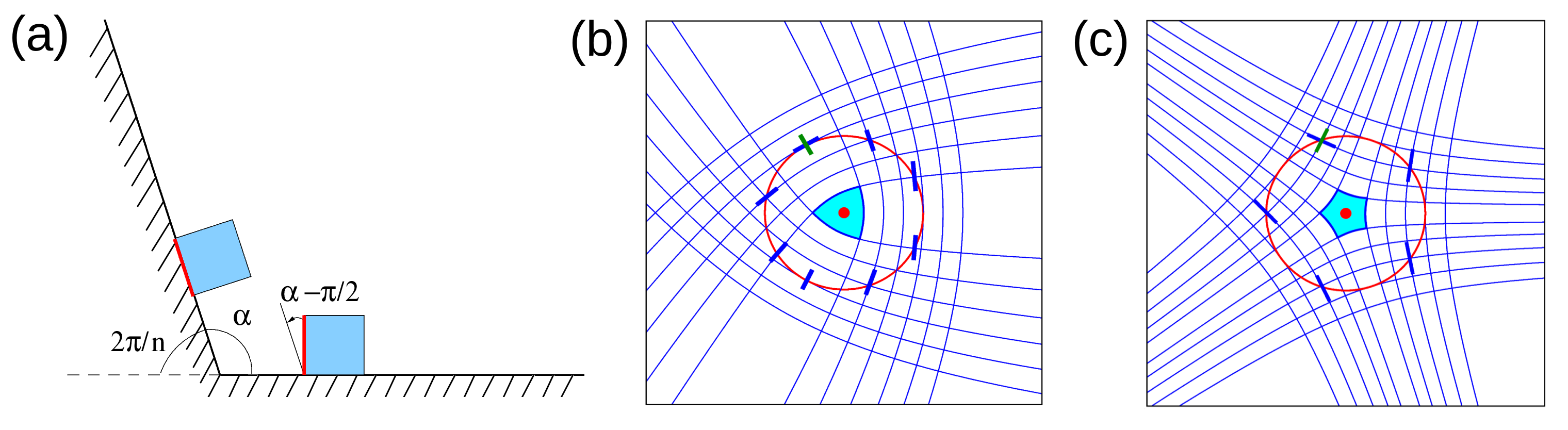}
\caption{(a) Schematic of square particles pertaining to perfect monolayers on 
consecutive sides of an $n$-sided polygon, representing the positions before and after 
a rotation by $\alpha-\pi/2$. (b) and (c) Curves of director orientations around 
free defects of winding numbers $+1/4$ (panel (b), charge $+1$) and $-1/4$ (panel (c), charge $-1$).}
\label{fig12}
\end{figure*}

Let $C$ denote the total number of defect configurations containing no more than
three free defects. This number depends on $n$, the number of sides of the confining polygon. We also define 
\begin{eqnarray}
{\cal N}=m_1+m_2+j_{1}+j_{-1}=n+j_{1}+j_{-1}
\end{eqnarray}
as the total number of defects (both pinned and free) of a given configuration, again under the constraint that there be at most three free defects. The value of ${\cal N}$ varies between a minimum and a maximum, given by
\begin{eqnarray}
{\cal N}_{\rm min}=\text{max}\left(n,2n-6\right),\hspace{0.2cm}\hbox{and}\hspace{0.2cm}{\cal N}_{\rm max}=n+3,
\end{eqnarray}
for $n\leq 9$. These bounds represent the smallest and largest possible total numbers of topological charges in the allowed configurations.
Table \ref{table4} summarizes the values of 
$C$, ${\cal N}_{\rm min}$ and ${\cal N}_{\rm max}$ as a function of $n$. 
The cases $n=3$ and $n=6$ are excluded from consideration, as these geometries  do not induce topological defects, as previously discussed. For $n\geq 10$ all possible configurations necessarily involve more than three free defects and are therefore beyond the scope of our analysis.

As the number of the sides of 
a polygon inscribed in a circle of fixed radius increases, the elastic energy contribution from the pinned defects, each carrying a positive charge, also increases, due 
to their mutual repulsion. To maintain the total topological charge at $+6$ an increasing number of negatively charged free defects must emerge to compensate for the growing excess 
of positive charge concentrated at the corners of the cavity.
However, configurations involving negative free defects tend to have reduced lifetimes, as previously discussed. For these reasons, we expect that, beyond a certain critical number of sides, likely equal to $n^*=10$ (see Table \ref{table4}), the system will transition from a polygonal to an effectively circular confinement. In this regime, the configuration consists of six free defects, each with charge $+1$, consistent with the known behavior in circular geometries. 

This conjecture could be tested in future experimental work planned in our laboratory. It is important to note that these
conclusions apply to cavities with dimensions comparable to those used in the present study. For larger cavities, the number of free defects may exceed three, potentially giving rise to a larger number of defects and a richer variety of defect configurations.

Finally, it is fruitful to consider other particle geometries. Triangular particles exhibit a unique characteristic: perfect monolayers formed on flat surfaces can adopt two distinct configurations: either one vertex or one side in contact with the surface. As a result, when confined in polygonal cavities, such granular fluids develop two types of pinned defects at the corners, with topological charges $+1$ or $+2$. 
In contrast, particle geometries that allow only a single stable configuration in contact with the cavity walls produce just one type of pinned defect. A paradigmatic case is a granular fluid composed of square particles. 
Following a director line from one wall to the next (assuming perfect anchoring conditions of squares at both walls) accumulates an interior rotation angle 
given by $\alpha-\pi/2=\pi/2-2\pi/n$, see Fig. \ref{fig12} (a). The total accumulated rotation includes both this interior rotation and the additional rotation at the vertex, $2\pi/n$), summing to $\pi/2$. The resulting winding number of the defect is then $k=(\pi/2)/(2\pi)=1/4$. Given the four-fold symmetry of the tetratic phase ($p=4$), the topological charge is $q=kp=+1$. This leads to a conservation law for the total topological charge within a polygonal cavity of $n$ sides, $n+j_1-j_{-1}=4$, where $j_1$ and $j_{-1}$ denote the number of free defects with charges $+1$ and $-1$, respectively (see panels (b) and (c) of Fig. \ref{fig12}). This constraint limits the the total number of free defects to 
$j_{1}+j_{-1}\leq 3$, under the conditions $m_2=0$ and $m_1=n$ (i.e. all corners are occupied by pinned defects of charge $+1$). Examples of allowable configurations include:
\begin{itemize}
\item $n=3$ (triangular cavity): there are two possibilities, (i) $j_1=1$ and $j_{-1}=0$, or (ii) $j_1=2$ and $j_{-1}=1$.
\item $n=4$ (square cavity): no free defects appear, as the square cavity is compatible with tetratic symmetry.
\item $n=5$ (pentagonal cavity): (i) $j_1=0$ and $j_{-1}=1$, or (ii) $j_1=1$ and $j_{-1}=2$.
\item $n=6$ (hexagonal cavity): only one configuration with $j_1=0$ and $j_{-1}=2$.
\item $n=7$ (heptagonal cavity): the only configuration has $j_1=0$ and $j_{-1}=3$.
\item $n=8$ (octogonal cavity): no configurations satisfy the constraint with three or fewer free charges.
\end{itemize}
This analysis clearly shows that particles with geometries that allow only a single anchoring configuration, such as squares, form monolayers that, when confined in polygonal 
cavities, present significantly fewer defect configurations compared to triangular particles.

\end{document}